# Мощные оптоэлектронные коммутаторы нано- и пикосекундного диапазона на основе высоковольтных кремниевых структур с *p-n*-переходами. III. Эффекты саморазогрева.

**А. С. Кюрегян**

ПАО "НПО "ЭНЕРГОМОДУЛЬ" , 109052, Москва, Россия

Впервые теоретически изучены эффекты саморазогрева оптоэлектронных коммутаторов на основе вертикальных высоковольтных структур с *p-n*-переходами (VPSS) при работе в высокочастотном режиме. Показано, что сильная зависимость коэффициента поглощения $\kappa(T)$ управляющего излучения от температуры $T$ является основным фактором, определяющим максимальную частоту коммутации $f_{\max}$ и соответствующую максимальную температуру кристалла $T_{\max}$, а также распределения температуры $T$ и плотности тока $j$ по площади VPSS. Двумерный анализ простейшей электротепловой модели VPSS, встроенного в двойную коаксиальную формирующую линию, показал, что увеличение частоты коммутации $f$ приводит к вытеснению тока на периферию прибора, где температура минимальна. Однако при частоте $f < f_{\max}$ и $T < T_{\max}$ распределения $T$ и $j$ по площади прибора остаются устойчивыми. Разумеется, величины $f_{\max}$ и $T_{\max}$ зависят от энергии управляющих импульсов излучения, импульсной коммутируемой мощности и условий теплоотвода. Для VPSS на основе непрямозонных полупроводников (Si, SiC) они изменяются в пределах 20-120 кГц и 120-160 °С, вполне достаточных для практического применения. Однако VPSS на основе прямозонных полупроводников (GaAs, InP) фактически не пригодны для работы в высокочастотных режимах из-за слишком резкой зависимости $\kappa(T)$.

В работах [1-3] были изложены теория и результаты численного моделирования изотермического переключения высоковольтных кремниевых структур с *p-n*-переходами (VPSS – Vertical Photoactivated Semiconductor Switch), управляемых пикосекундными лазерными импульсами, которые позволили получить соотношения между параметрами коммутаторов и характеристиками процесса переключения. Эти результаты применимы для анализа работы и проектирования VPSS на низких частотах $f$, так как за один цикл коммутации в оптимальных режимах рассеивается энергия с плотностью $w_D \sim 1$ мДж/см$^2$ [3] и средний по объему перегрев не превосходит 0,01 К. Однако при $f \geq 50$ кГц VPSS с освещаемой площадью $S_{ph} \sim 1$ см$^2$ рассеивает среднюю мощность $w_D S_{ph} f \geq 50$ Вт и его квазистационарный саморазогрев может превышать 100 К, так как в СВЧ тракте (куда должен быть встроен VPSS) трудно обеспечить теплоотвод с сопротивлением $R_T \leq 2$ К/Вт. Поэтому для решения практических задач проектирования VPSS необходимо ясное понимание эффектов их саморазогрева, которые обладают нетривиальной особенностью.

В обычных высоковольтных приборах повышение температуры $T$ приводит к изменению подвижностей $\mu_{e,h}(T)$ и времён жизни $\tau_{e,h}(T)$ электронов (*e*) и дырок (*h*), а также к уменьшению контактной разности потенциалов *p-n*-переходов $V_C(T)$. Так как обычно длительность формируемого VPSS импульса $t_R \ll \tau_{e,h}$ и меньше времени $t_{sc}$ начала возникновения области пространственного заряда (см. [2]), а падение напряжения на VPSS $U \gg V_C$, то последними двумя эффектами можно пренебречь. Но вместо них появляется новый эффект разогрева – увеличение коэффициента поглощения управляющего излучения $\kappa(T)$, анализу которого посвящена настоящая работа. Она является продолжением [2-3], так что все обозначения, конструкция VPSS, режимы освещения и коммутации остались теми же, а все результаты приведены для случая $S_{ph} = 0.5$ см$^2$, площади прибора $S_D = 2S_{ph}$, энергии управляющих импульсов света $W_{ph} = 50$ мкДж, длительности им-пульсов тока $t_R = 10$ нс и, если это специально не оговорено, температуры окружающей среды $T_{ext} = 20\,^{\circ}\mathrm{C}$.

Суть дела состоит в том, что энергия $W_D = w_D S_{ph}$, рассеиваемая прибором за время $t_R$, немонотонно зависит от $\kappa$ [3]. При $kd \ll 1$ увеличение $\kappa$ приводит к росту количества порождаемых светом носителей заряда, распределенных почти однородно по толщине кристалла. Поэтому эффективное сопротивление VPSS в проводящем состоянии и, следовательно, $W_D$ уменьшаются. При $kd > 1$ увеличение $\kappa$ приводит ко все более медленному росту количества носителей заряда, но теперь они сосредоточены в слое с уменьшающейся толщиной $\kappa^{-1} < d$ вблизи освещаемой поверхности, поэтому сопротивление VPSS и $W_D$ возрастают при увеличении $\kappa$. Минимум функции $W_D(\kappa)$ достигается при $\kappa \approx 2/d$ (см. Рис. 8 в [3]), так что оптимальная длина волны света $\lambda$ попадает на край основной полосы поглощения, где $\kappa$ очень сильно зависит и от $\lambda$, и от $T$. Примеры зависимостей $\kappa(T)$ для четырех полупроводников по данным работ [4-9] приведены на Рис. 1. Все они хорошо аппроксимируются степенной функцией

$$\kappa(T) = \kappa_0 \left(T/T_0\right)^{\gamma}, \qquad (1)$$

предложенной авторами работы [5] для поглощения света с $\lambda = 1064$ нм в Si. С ростом $\lambda$ показатели степени $\gamma$ слабо увеличиваются, но почти пятикратное различие между их значениями для прямозонных и непрямозонных материалов сохраняется.

Вследствие этого зависимость $W_D(T)$ также оказывается немонотонной. Пример такой зависимости для случая однородных распределений тока и температуры по площади кремниевого VPSS, освещаемого светом с $\lambda = 1064$ нм, приведен на Рис. 2, кривая A. Следует отметить, что температурная зависимость подвижностей очень слабо влияет на вид функции $W_D(T)$. На это указывают результаты расчетов (кривая B на Рис. 2) при различных температурах с учетом реальной зависимости $\mu_{e,h}(T)$ и постоянном значении $\kappa = 35$ см$^{-1}$, соответствующем температуре 100 °C.

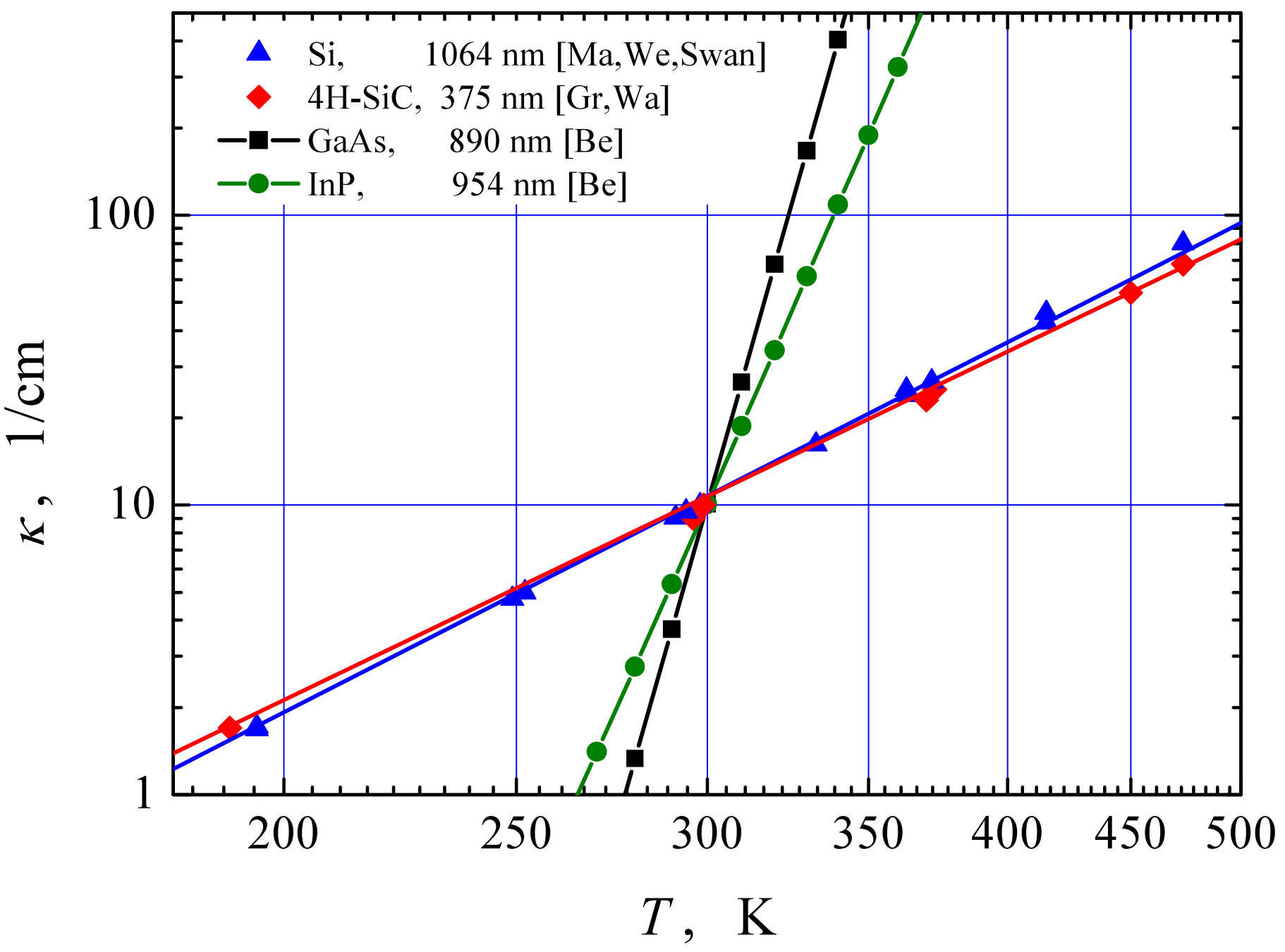

Рис. 1. Температурные зависимости коэффициента поглощения в Si (треугольники, $\lambda = 1064$ нм [4-6]), 4H-SiC (ромбы, $\lambda = 375$ нм [7,8]), GaAs (квадраты, $\lambda = 890$ нм [9]) и InP (кружки, $\lambda = 954$ нм [9]) излучения с длинами волн $\lambda$, для которых $\kappa(300\ \mathrm{K}) \approx 10\ \mathrm{cm}^{-1}$. Линии – аппроксимации по формуле (1) с показателями степени $\gamma = 4.25$ для Si, $\gamma = 4.0$ для 4H-SiC, $\gamma = 29.0$ для GaAs и $\gamma = 19.0$ для InP.

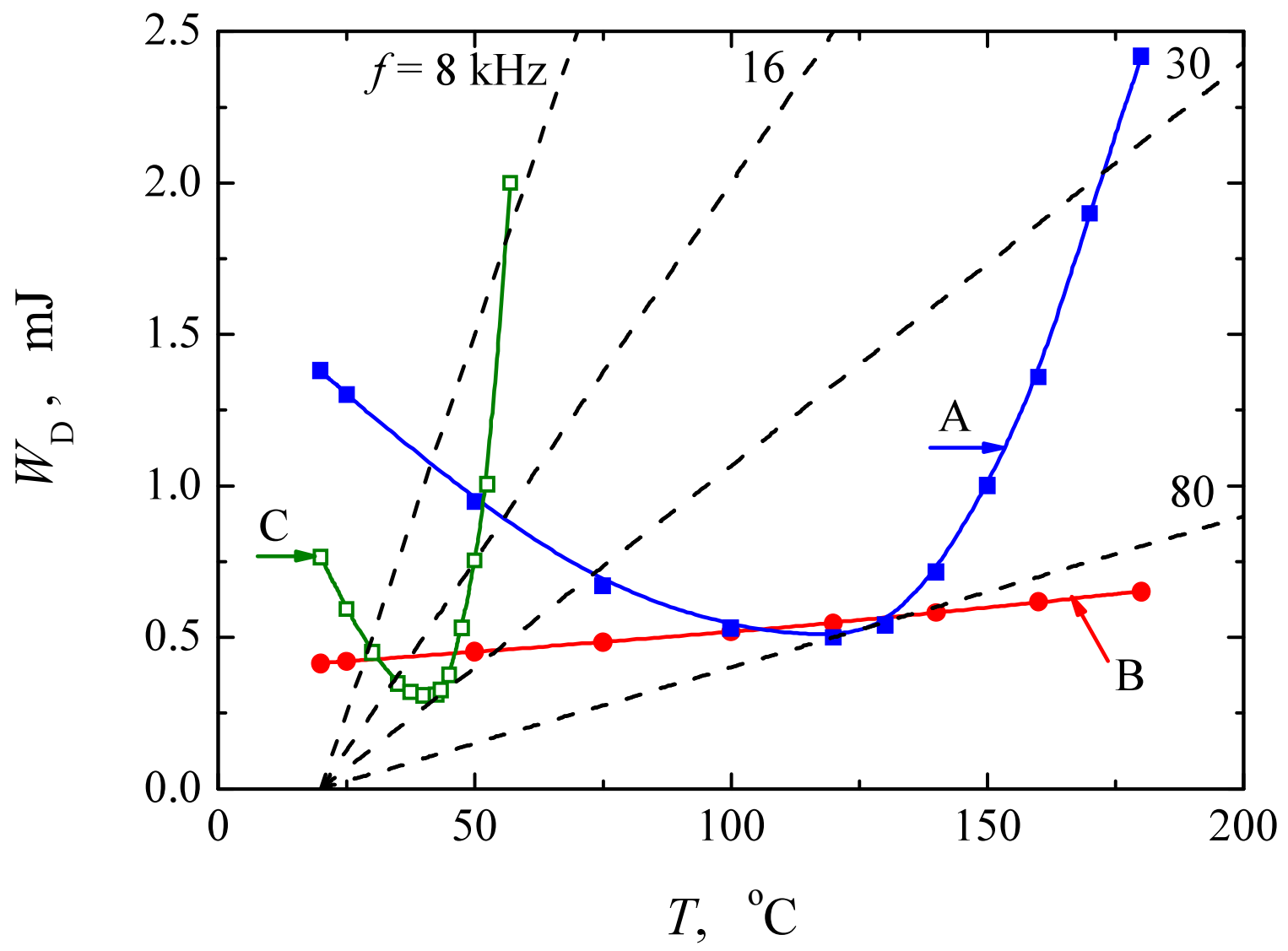

Рис. 2. Сплошные линии - зависимости энергии $W_{\rm D}$ от температуры (см. пояснения в тексте). Штриховые линии – энергия $W_T = (T - T_{ext})/R_T f$, отводимая за время $1/f$ системой охлаждения с тепловым сопротивлением $R_T = 2.5$ K/W при различной частоте повторения импульсов $f$. Символы – результаты численного моделирования в одномерном приближении по данным работы [3].

При работе в высокочастотном режиме температура очень слабо изменяется за время $1/f$, по толщине $d$ кристалла и при изменении радиальной координаты $\rho$ на размер элементарной ячейки VPSS, который должен быть меньше или порядка $d$. Поэтому для анализа эффектов саморазогрева приборов с площадью $S_D >> d^2$ целесообразно использовать температуру, усредненную по времени $1/f$ и объему $d^3$. Далее для упрощения формул мы будем обозначать эту «среднюю» температуру тем же символом $T$. После включения прибора $T$ увеличивается с постоянной времени $\tau_D >> 1/f$ до тех пор, пока средняя по времени рассеиваемая мощность $P_D = W_D(T) f$ не сравняется с мощностью $(T - T_{ext})/R_T$, отводимой системой охлаждения однородно по площади. Через время $t >> \tau_D$ возникает стационарное состояние с температурой прибора $T_{st}$, которая является минимальным корнем уравнения

$$T_{st} = T_{ext} + W_D(T_{st}) R_T f, \qquad (2)$$

где $T_{ext}$ - температура окружающей среды. Из Рис. 2 следует, что его решение существует только при частоте, меньшей некоторого значения $f_{\max}$, которое можно вычислить приближенно по формуле

$$f_{\max} \approx \frac{T_{\max} - T_{ext}}{W_D(T_{\max}) R_T}, \qquad (3)$$

где $T_{\max} \approx T_W$, а $T_W$ - температура, при которой энергия $W_{\rm D}$ минимальна. В рассматриваемом конкретном случае $T_{\max} = 120\,^\circ$C, $W_D(T_{\max}) \approx 0.5$ мДж и $f_{\max} \approx 80$ кГц для кремниевого VPSS. Для приближенного вычисления температуры $T_{\max}$ и энергии $W_D(T_{\max})$, которые зависят от параметров прибора и управляющего импульса света, можно использовать следующее соотношение между плотностью импульса тока $j$ в освещаемых областях и падением напряжения $U_{on}$ на приборе в проводящем состоянии:

$$j(T, U_{on}) \approx j_0 \Psi\left[\kappa(T) d, U_{on}/\bar{E}d\right], \qquad \Psi(k,u) = k\frac{\exp(ku) - 1}{\exp\left[k(u+1)\right] - 1}, \qquad (4)$$

где $j_0 = \frac{W_{\rm ph}}{\hbar\omega}\frac{2q\bar{v}}{S_D d}$, $\hbar\omega$ - энергия квантов света, $q$ - элементарный заряд. При выводе (4), как и аналогичной формулы (7) из работы [2], мы предполагали выполнение неравенства $t_R < t_{sc}$, пренебрегали током смещения в проводящем состоянии прибора и резистивным падением напряжения вдоль электродов, обусловленным растеканием тока. Кроме этого, не учитывались отражение света от тыльного контакта и различие подвижностей электронов и дырок. Эти два упрощения приводят к некоторой количественной погрешности, но зато позволяют на качественном уровне учесть эффект насыщения зависимости дрейфовых скоростей электронов и дырок от напряженности поля $E$ в соответствии с обычной формулой $v(E) = \bar{v}E/(E+\bar{E})$, который становится важным при очень малых и очень больших значениях $kd$. Графики зависимостей $j(T,U_{on})$ приведены на Рис. 3.

Энергия, рассеиваемая прибором за один импульс длительностью $t_R$, равна

$$W_D = W_C + t_R I U_{on}, \qquad (5)$$

где $W_C = C_D U_0^2/2 \approx 0.24$ mJ - энергия, накопленная в барьерной емкости прибора $C_D$ и рассеиваемая им в процессе переключения [3], $I$ - полный ток через прибор, равный

$$I = (U_0 - U_{on})/R, \qquad (6)$$

$R = 2Z$ для одинарной и $R = Z$ для двойной формирующей линии с волновым сопротивлением $Z$, согласованной с активной нагрузкой. Далее, как и в [2,3], мы будем считать, что $R = 10$ Ом. Если температура и ток однородны по площади VPSS, то $I = S_{ph} j(T,U_{on})$ и совместное использование (4)-(6) позволяет просто найти $T_{\max}$ и $W_D(T_{\max})$ и вычислить $f_{\max}$ по формуле (3).

Однако однородный по площади теплоотвод практически невозможно реализовать, поскольку высокая скорость переключения может быть «утилизирована» надлежащим образом, если только сам VPSS является составной частью формирующей линии. Возможный вариант такой конструкции на основе двойной коаксиальной формирующей линии изображен на Рис. 4. Тепло отводится от VPSS к радиатору через полупрозрачный молибденовый электрод, обеспечивающий квазиоднородную засветку, и кольцевой керамический изолятор. Радиатор обеспечивает теплоотвод с тепловым сопротивлением $r_T$ от внешней образующей поверхности керамического изолятора к окружающей среде. Теплоотвод через дисковую пружину, соединяющую VPSS с внутренним проводником формирующей линии, пренебрежимо мал. Ясно, что в этом случае температура кристалла, и, значит, плотность тока, не могут быть однородными по площади.

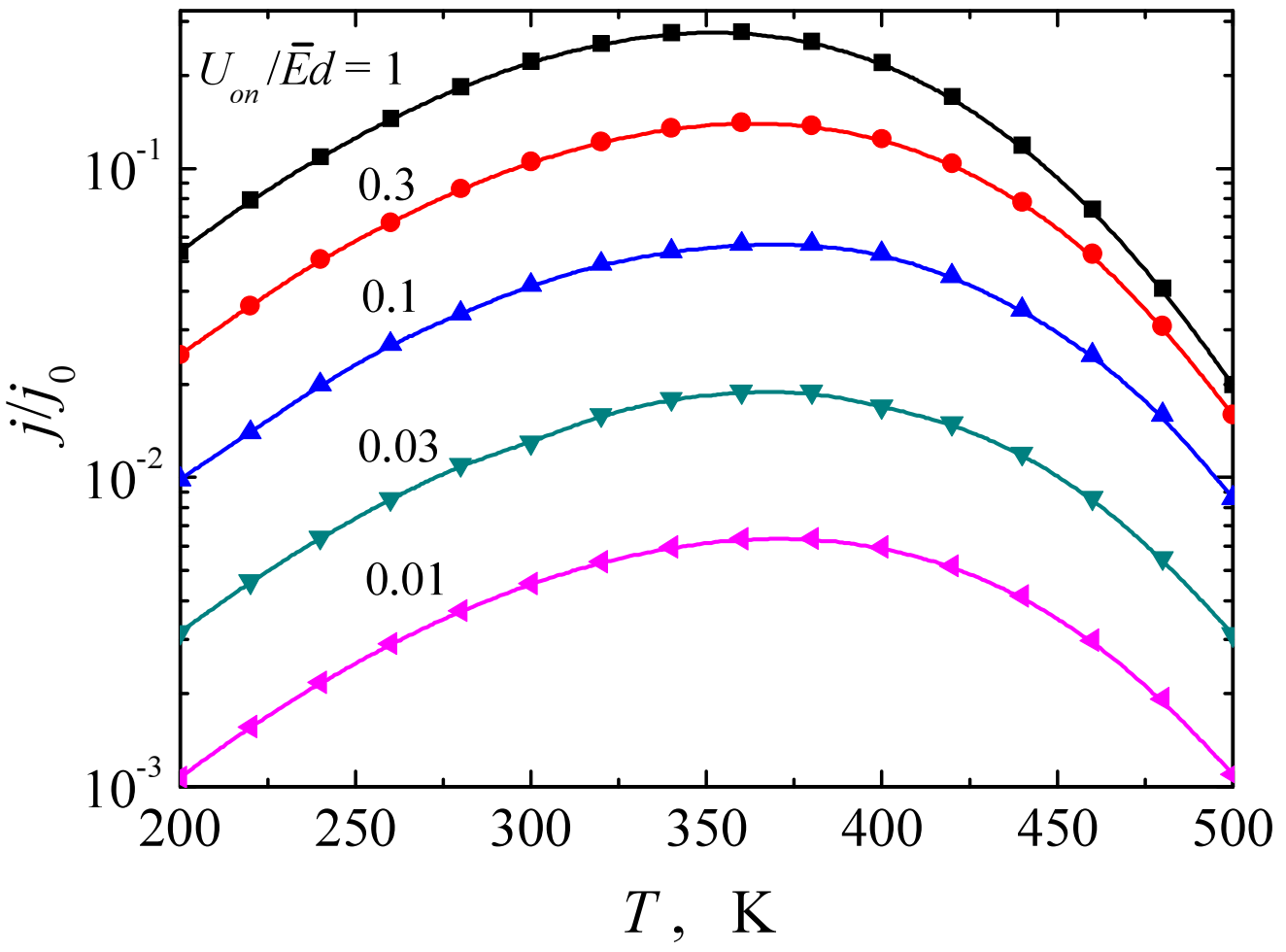

Рис. 3. Зависимости $j(T,U_{on})$, рассчитанные по формуле (4) для «кремниевой» зависимости $\kappa(T)$.

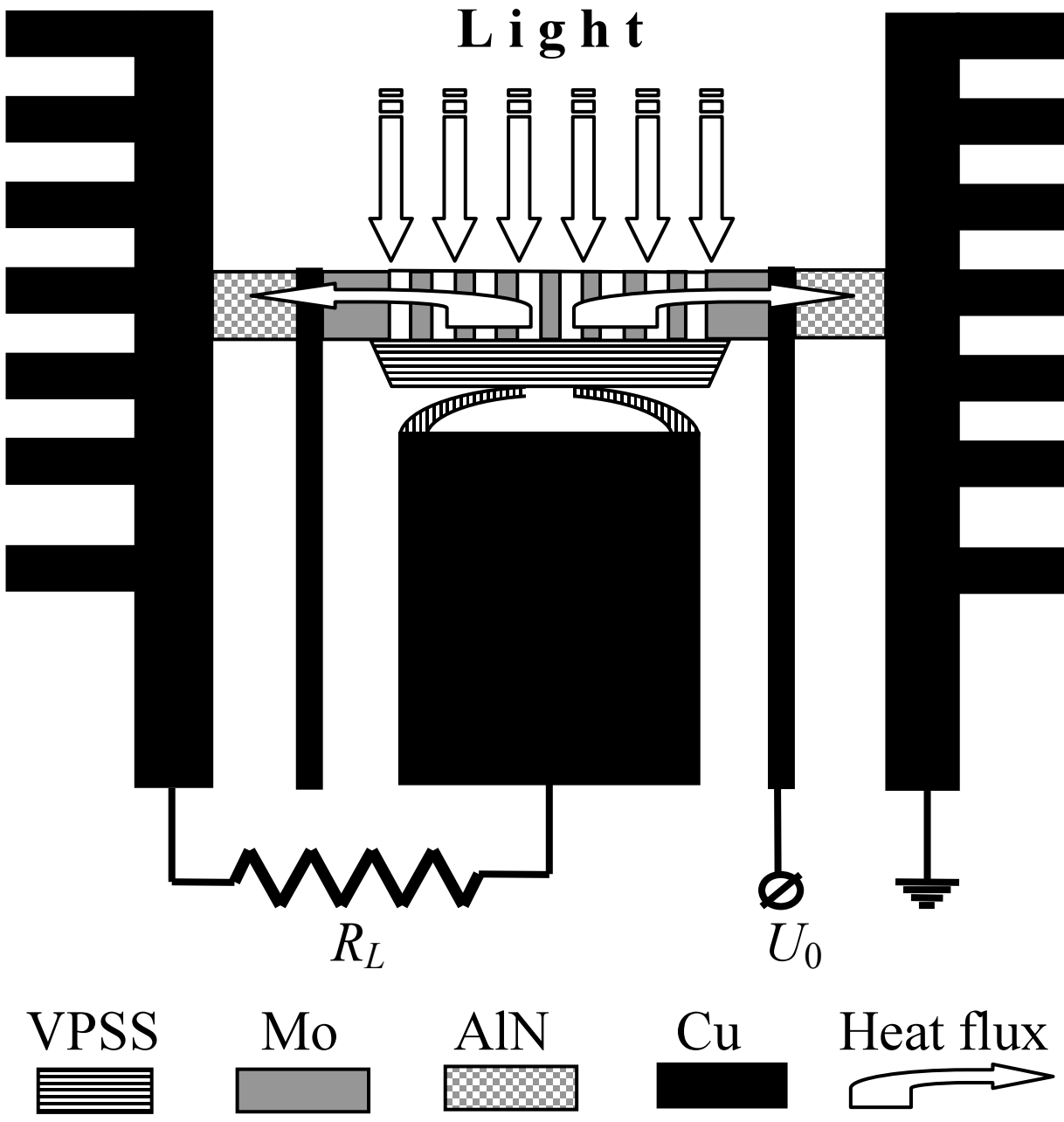

Рис. 4. Схематичное изображение вертикального оптоэлектронного коммутатора, присоединенного к торцу двойной коаксиальной формирующей линии с волновым сопротивлением $Z$, согласованной с активной нагрузкой $R_L = 2Z$ [10].

Для вычисления распределений $T(\rho)$ и $j(\rho)$ по радиусу VPSS необходимо решить уравнение теплопроводности с учетом того, что средняя[1] плотность мощности $p_D$, рассеиваемой прибором, зависит от температуры:

$$p_D(T,U_{on}) \approx \left\{W_C + t_R j_0 S_{ph} U_{on} \Psi\left[\kappa(T)d, U_{on}/\bar{E}d\right]\right\} f/S_D \,, \tag{7}$$

а полный ток определяется формулой

$$I = \frac{S_{ph}}{S_D} \int_{S_D} j(T,U_{on})\, dS \,. \tag{8}$$

Мы решали эту задачу численно для конструкции, изображенной на Рис. 4, используя соотношения (4)-(8) и считая, что радиус молибденового электрода $\rho_{\mathrm{Mo}} = 7$ мм, внешний радиус керамического изолятора $\rho_{\mathrm{AlN}} = 9$ мм, а толщина эти двух деталей равна 1 мм. Результаты расчетов приведены на Рис. 5-7.

Как и следовало ожидать, температура в центре прибора $T(0)$ всегда максимальна (Рис. 5a). При малых частотах плотность тока также максимальна в центре и очень слабо уменьшается к периферии прибора. Однако при $f > 50$ кГц температура $T(0)$ превышает значение 375 К, при котором достигается максимум функции $j(T)$ (см. Рис. 3), тогда как температура на периферии $T(\rho_D)$ остается меньше 375 К. Вследствие этого ток вытесняется на периферию прибора (Рис. 5b), где величина коэффициента поглощения ближе всего к оптимальному значению $\kappa(T) \approx 2/d$, условия теплоотвода улучшаются и поэтому эффективное тепловое сопротивление $R_T = \left[T(0) - T_{ext}\right]/P_D$ уменьшается, как это изображено на Рис. 6. Кроме того при $f > 50$ кГц падающая зависимость $U_{on}(f)$ сменяется на быстро нарастающую. Это приводит к сверхлинейному росту рассеиваемой мощности $P_D(f) = U_{on} I$, так как при нормальной работе VPSS $U_{on} << U_0$ и полный ток $I \approx U_0/R$, то есть практически не зависит от $f$ (см. формулу (6)). С ростом $f$ мощность $P_D$ нарастает быст-рее, чем падает тепловое сопротивление $R_T$ (см. Рис. 6), поэтому при $f > f_{\max}$ условия существования стационарного состояния перестают выполняться и наступает тепловой пробой.

[1] Усреднение плотности рассеиваемой мощности по времени и координате производилось так же, как усреднение температуры.

Разумеется, $f_{\max}$ зависит от параметров VPSS, управляющих импульсов света, формирующей линии и условий теплоотвода. В качестве примера на Рис. 7 приведены зависимости $f_{\max}$ от $r_T$ и $T_{ext}$. Как и следовало ожидать, $f_{\max}$ увеличивается при уменьшении $r_T$ и $T_{ext}$, но распределения температуры $T(\rho)$ и плотности тока $j(\rho)$ по радиусу $\rho$ кристалла становятся все более неоднородными. Однако эти распределения остаются устойчивыми вследствие того, что при $T > T_W$ плотность тока уменьшается с ростом температуры кристалла.

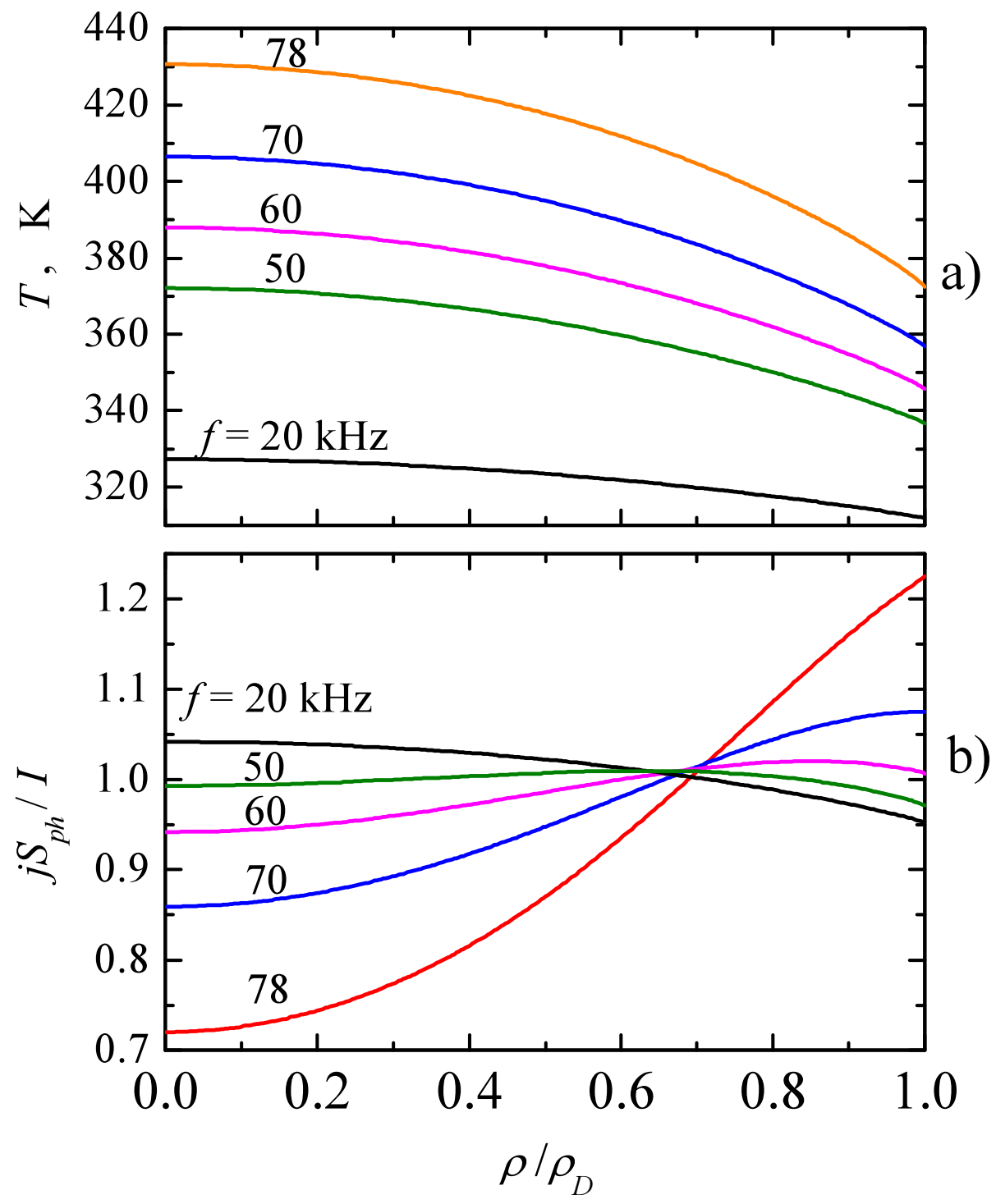

Рис. 5. Радиальные распределения температуры кристалла (a) и нормированной плотности тока (b) при $r_T = 1$ k/W и различных частотах повторения импульсов $f$.

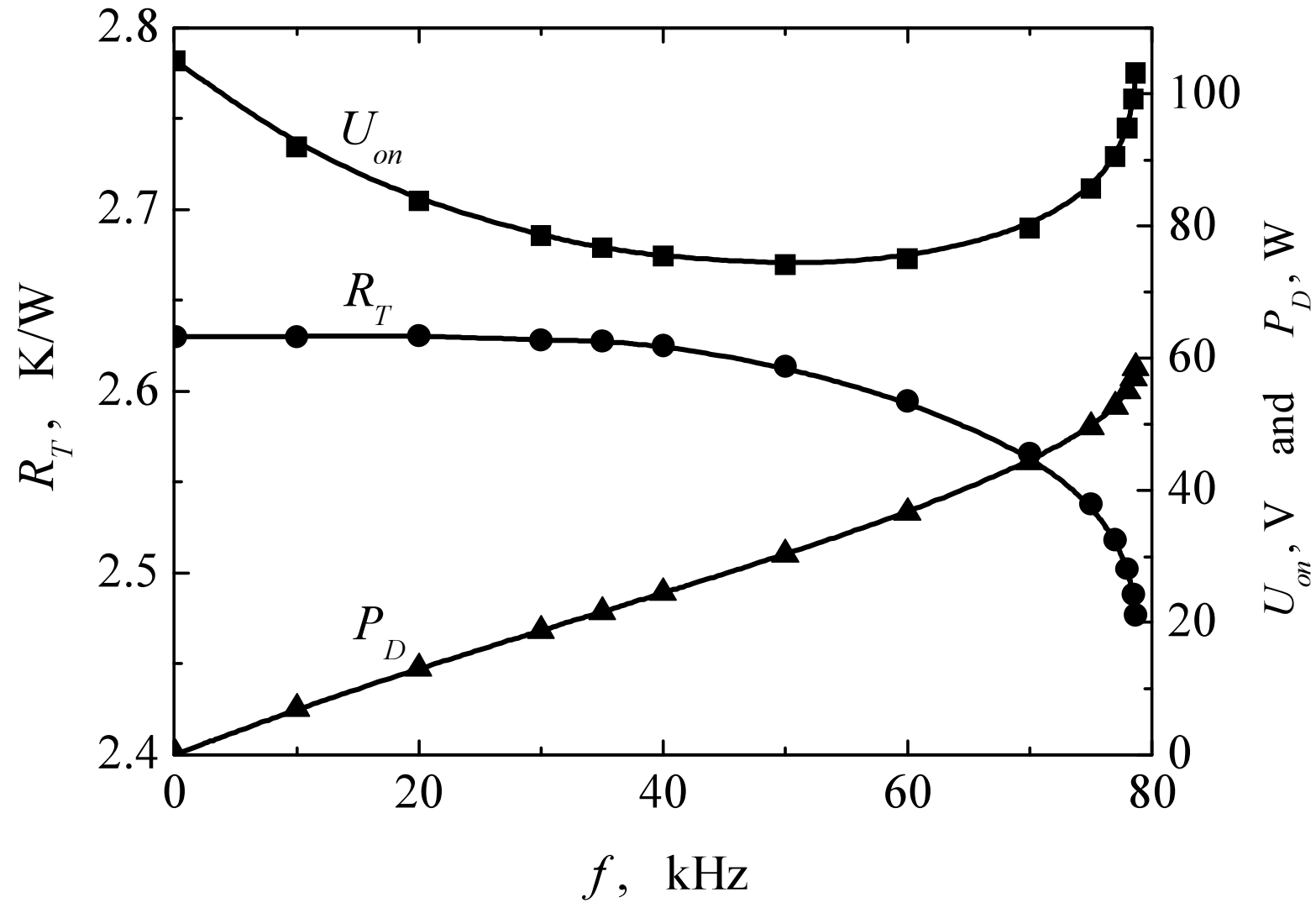

Рис. 6. Зависимости эффективного теплового сопротивления $R_T$ (кружки), падения напряжения на VPSS $U_{on}$ (квадраты) и средней рассеиваемой мощности $P_D$ (треугольники) от частоты при $r_T = 1$ k/W.

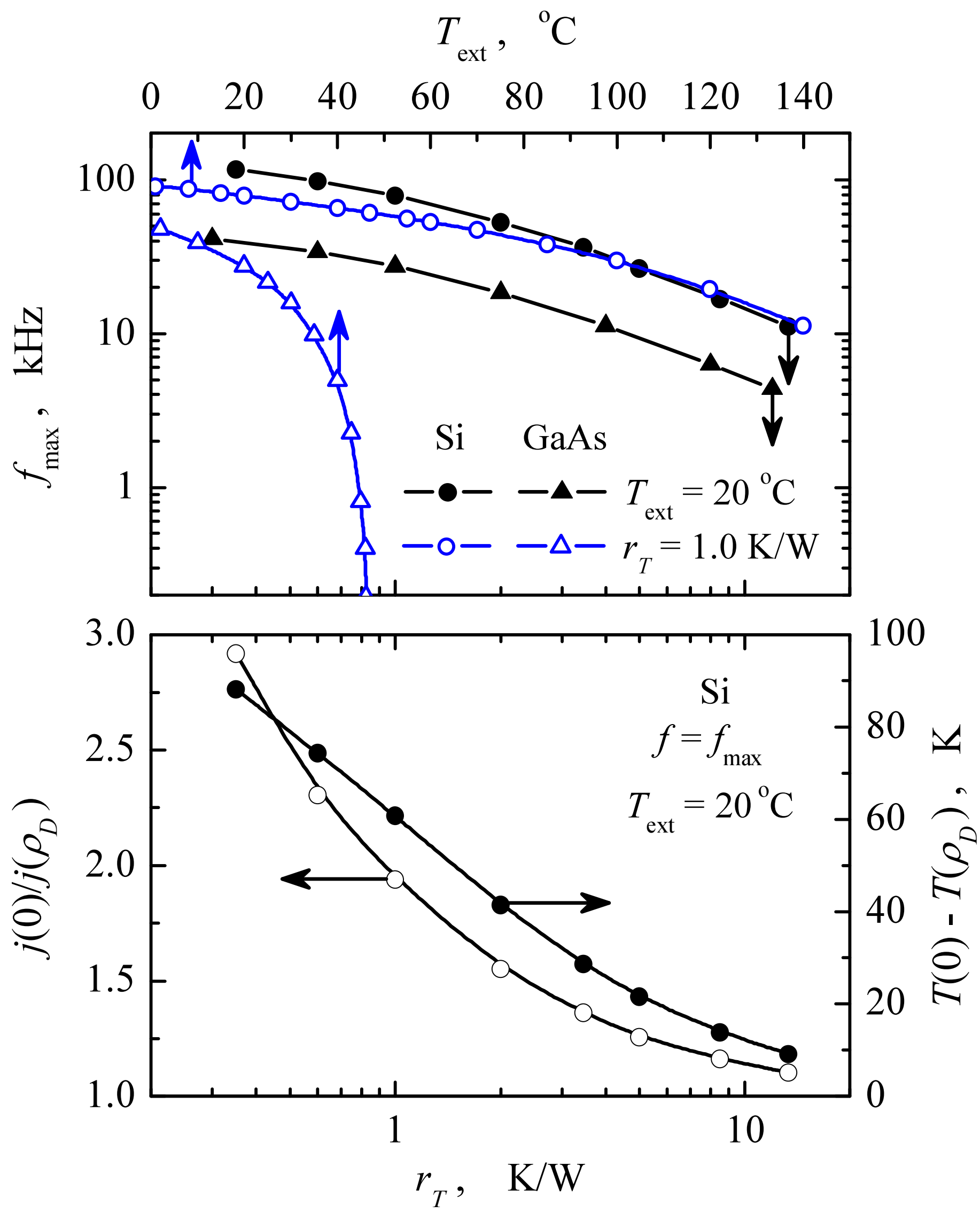

Рис. 7. Вверху: зависимости максимальной частоты повторения импульсов $f_{\max}$ от $r_T$ при $T_{ext} = 20\,^0\mathrm{C}$ (темные символы) и от $T_{ext}$ при $r_T = 1$ k/W (светлые символы) для кремниевого (кружки) и арсенид-галлиевого (треугольники) VPSS. Внизу: зависимости отношения плотностей тока (светлые символы) и разности температур (темные символы) в центре и на периферии кремниевого VPSS от $r_T$ при $f = f_{\max}$ и $T_{ext} = 20\,^0\mathrm{C}$.

Видно также, что кремниевые VPSS могут работать в высокочастотном режиме при очень высоких температурах окружающей среды. Это качество, обусловленное тем, что зависимость $\kappa(T)$ в Si не слишком резкая ($\gamma = 4.25$), присуще также и VPSS на основе карбида кремния ($\gamma = 4.0$). В этом отношении VPSS на основе прямозонных полупроводников типа GaAs и InP значительно хуже. Во-первых, и при $T_{ext} = 20\,^0\mathrm{C}$ их максимальная рабочая частота оказывается примерно в 3 раза меньше несмотря на то, что суммарная подвижность электронов и дырок в GaAs в 4 раза больше, чем в Si. Во-вторых, $f_{\max} \to 0$ при $T_{ext} \to 45\,^0\mathrm{C}$, так что при бóльших температурах арсенид-галлиевые VPSS вообще не может функционировать в частотном режиме. Причиной этого является гораздо бóльшее значение $\gamma$ (см. Рис. 1) и соответствующее очень резкое возрастание $W_D(T)$ при отклонении температуры кристалла от значения $T_W \approx 40\,^\circ\mathrm{C}$, изображенное кривой C на Рис. 2.

Работа выполнена при поддержке РФФИ (грант № 16-08-01292)

## Литература